\documentclass{elsart}
\usepackage{natbib}
\begin{document}
\runauthor{H. J. V\"olk and A. M. Atoyan}
\begin{frontmatter}
\title{Clusters of Galaxies: magnetic fields and nonthermal emission}
\author{H. J. V\"olk$^{\rm a}$ and   A. M. Atoyan$^{\rm a, b}$ }
\address[MPIK]{Max-Planck-Institut f\"ur Kernphysik, D-69029 Heidelberg,
 Germany}
\address[YER]{Yerevan Physics Institute, 375036 Yerevan, Armenia}

\begin{abstract}
The nonthermal particle content of galaxy clusters should in part have a
cosmological component generated during the early starburst phase of the
member galaxies. This is reviewed in the framework of a simple cluster
formation model suggested previously. It implies a nonthermal energy
fraction of about 10 percent for the Intracluster gas. We also propose a
mechanism for the early generation of Intracluster magnetic fields in
terms of Galactic Winds. It results in typical field strengths of
$10^{-7}$ \, Gauss. Such comparatively weak fields are consistent with an
inverse Compton origin of the excess EUV and hard X-ray emission of the
Coma cluster, given the radio synchrotron emission. The required relativistic
electrons must have been accelerated rather recently, less than a few
billion years ago, presumably in cluster accretion shocks. This is in contrast
to the hadronic nonthermal component which accumulates on cosmological
time scales, and whose $\pi^0$-decay TeV $\gamma$-ray emission is expected
to be larger, or of the same order as the inverse Compton TeV emission. 
This $\gamma$-radiation characterizes the energetic history of
cluster formation and should be observable with future arrays of imaging
atmospheric Cherenkov telescopes.                        
\end{abstract}
\end{frontmatter}

\section{Introduction}

Clusters of galaxies are the largest gravitationally bound structures in
the Universe and may confine a representative fraction of its mass.
Therefore the study of their dynamical properties and radiation content 
should allow, amongst other things, conclusions on the
relative amounts of baryonic and nonbaryonic matter in cosmology (e.g.
White and Fabian, 1995, and references therein).

Another basic characteristic, also due to confinement, is the ratio
of thermal to nonthermal energy in these objects. To a significant extent
that ratio is established during the epoch of galaxy formation and thus
preserves the energetic history of cluster formation. We shall review this
topic here. The confinement of nonthermal particle components is
intimately related to the existence of strong and chaotic 
magnetic fields in the intracluster medium (ICM),
and we shall propose a mechanism for their early
generation. This is followed by a discussion of the present-day
nonthermal radiation from clusters in various wavelength ranges, 
in particular at very high $\gamma$-ray energies.

{\it Rich Clusters}

Rich clusters, i.e. conglomerates with typically more than 100 member
galaxies, have typical radii $R_{\rm cl} \sim$ few Mpc and baryonic
masses $M_{\rm cl} \sim 10^{14} \,{\rm to}\, 10^{15}\,{\rm M}_{\odot}$. 
The closest large cluster is
the Virgo cluster at a distance of $d \sim 20 \, {\rm Mpc}$; all the others
are at a distance $d$ of hundred Mpc and beyond. Examples for bright and
relatively nearby clusters are the Perseus and the Coma clusters with $d
\simeq 100\,\rm Mpc$. The Perseus cluster is the brightest cluster in soft
X-rays (see Fig. 1). 
The large X-ray luminosity is due to the very hot ($T \sim 10^7 \,\rm to \,
10^8\,$K), massive ($M_{\rm gas} \sim {\rm few} \times \sum M_{\rm gal}$), and 
metal-rich
($[{\rm Fe}]_{\rm cl} \simeq 0.35 [{\rm Fe}]_{\odot}$) ICM gas
(e.g. B\"ohringer, 1996).

Apart from their primary cosmological interest, galaxy clusters also serve
as extragalactic distance poles (Sunyaev-Zeldovich effect) and as
gravitational telescopes for still more distant objects. This makes them 
important "instruments" for observational astronomy.

\begin{figure}
\vspace{6cm}
\includegraphics{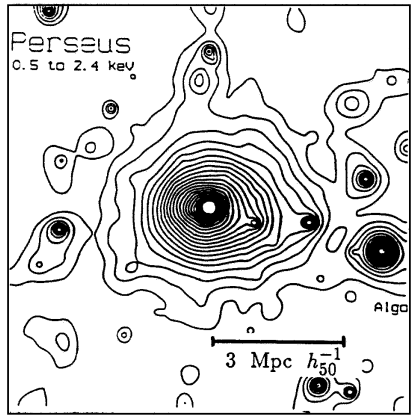}
\caption{
Large scale contour plot of an X-ray image of the Perseus cluster
from the ROSAT all sky survey (courtesy H. B\"ohringer).
 }
\end{figure}

{\it Cluster Formation} 

Most rich clusters - apart from objects still in an early formation
phase, like the Virgo cluster - are characterized by a predominance of
early type galaxies, i.e. elliptical (E) and lenticular (S0) galaxies
which have little interstellar gas. Gas-rich spiral and irregular galaxies
represent a minority. This situation is opposite to the one in the field,
the space outside clusters, where the fraction of spiral galaxies is at
least 75 percent. For rich clusters the metallicity of the ICM gas, for
instance in the form of the fractional ICM iron mass, is also correlated
with the optical luminosity in E and S0 galaxies (Arnaud et al., 1992).
The correlation supports the qualitative view that early starbursts due to
galaxy-galaxy interactions of protospirals have produced a large number of
Supernovae (SNe) that heated the originally present interstellar gas and
generated violent Galactic Winds, leaving gas-poor E and S0 galaxies
behind. This mass loss should have led ultimately to the observed strong
chemical enrichment of the ICM gas. We also conjecture that the radiation 
and the winds from these early galaxy mergers strongly heated the remaining
primordial ICM gas and thus prevented further galaxy formation.
This is perhaps the physical explanation for the observed inefficiency of
galaxy formation which manifests itself in the remarkable preponderance of
diffuse ICM gas mass over stellar mass in clusters of galaxies.

A quantitative discussion of the dynamical prerequisites for Galactic
Winds and the total number of SNe in clusters is contained in the paper by
V\"olk et al. (1996, hereafter referred to as
Paper I) which we shall summarize below. 

The total number of SNe since galaxy formation in the cluster, roughly a
Hubble time $T_{\rm H} \simeq 1.5 \times 10^{10} {\rm yr}$ ago, is given
by $$N_{\rm SN}=\int_{-T_{\rm H}}^{0}dt \times
\nu_{\rm SN}(t)=\frac{0.35 \,{[Fe]}_{\odot} \times M_{\rm cl}}{\delta
M_{Fe}}\;,$$
where $\delta M_{Fe}$ is the amount of iron produced per event. In such
starbursts we dominantly expect core collapse SNII from massive progenitor
stars to occur, with $\delta M_{Fe}\simeq 0.1 M_{\odot}$ on average. For
the Perseus cluster this implies 
$N_{\rm SN}^{\rm Perseus} \sim 3 \times 10^{12}$.
The corresponding total energy input into the interstellar medium is
$N_{\rm SN}E_{\rm SN} \sim 3 \times 10^{63} E_{51}\,\rm erg$,
where $E_{51}=10^{51} \,\rm erg$ is the average hydrodynamic energy
release per SN.

Assuming the early starbursts to occur at a typical redshift of $z\sim 2$ 
due to the merging of protospirals in the overdense protocluster
environment (Steinmetz, 1993), with a duration of 
$T_{\rm SB} \leq 10^9 \,\rm yr$, we obtain
$$
\frac{(N_{\rm SN}^{\rm Perseus}/N_{\rm gal}^{\rm Perseus})}{T_{\rm SB}}
\geq 100 \times \nu_{\rm SN}^{\rm Milky\, Way} .
$$
Here $N_{\rm gal}^{\rm Perseus} \!\simeq \! 500$ denotes the number of galaxies 
in the
cluster. As an example we can compare to the archetypical contemporary
starburst galaxy $M82$. It has a current SN rate $\nu_{\rm SN}^{M82} \sim
10 \times \nu_{\rm SN}^{\rm Milky\, Way}$, a wind 
velocity $v_{\rm wind} \sim 2300\, {\rm
km/sec}$, and a mass-loss rate of $\dot{M} \sim 0.8 M_{\odot}/{\rm yr}$
(Breitschwerdt, 1994). The starburst nucleus of $M82$ is characterized
by the following values for the interstellar gas temperature $T$, gas
density $n$, and thermal gas pressure $p$ : $T_{\rm base} \sim 10^8 \,{\rm K}$,
$n_{\rm base} \sim 0.3\, {\rm cm}^{-3}$, and 
$p_{\rm gas}/k_{B} \sim 10^7\, {\rm K \,cm^{-3}}$ (Schaaf et al.
1989). Since the thermal ICM gas pressure in the Perseus cluster is
$p_{\rm cl}^{\rm Perseus}/k_{B} \sim 10^4 \,{\rm K}\,
{\rm cm}^{-3}$, it is clear 
that an object like M82 could readily drive a wind even against the 
{\it present-day} ICM
pressure. At the galaxy formation epoch the ICM pressure was certainly much
smaller than this value.
  
In the expanding wind flow the SN-heated gas cools adiabatically to quite
small temperatures. However it is reheated in the termination
shock, where the ram pressure of the wind adjusts to the ICM pressure.
Beyond this point the ejected galactic gas is mixed with the unprocessed
ICM gas. 

{\it Particle Acceleration}

Cluster formation also implies the production of a strong nonthermal
component of relativistic particles. They will be accelerated during the
early phase - and possibly also in later events - and confined in the
turbulent ICM medium. The confinement time generally exceeds the cluster
lifetime. Thus the energy spectrum of the energetic particles is the same
as that generated by their sources. In Cosmic Ray parlance these are {\it
cosmological} Cosmic Rays (CRs).

During the early starburst particle acceleration will occur initially at
the outer shocks of the Supernova Remnants (SNRs). However, like the
thermal gas, this first generation of nonthermal particles will loose
its energy almost completely by adiabatic cooling in the ensuing
Galactic Wind. 

\begin{figure}
\vspace{6cm}
\includegraphics{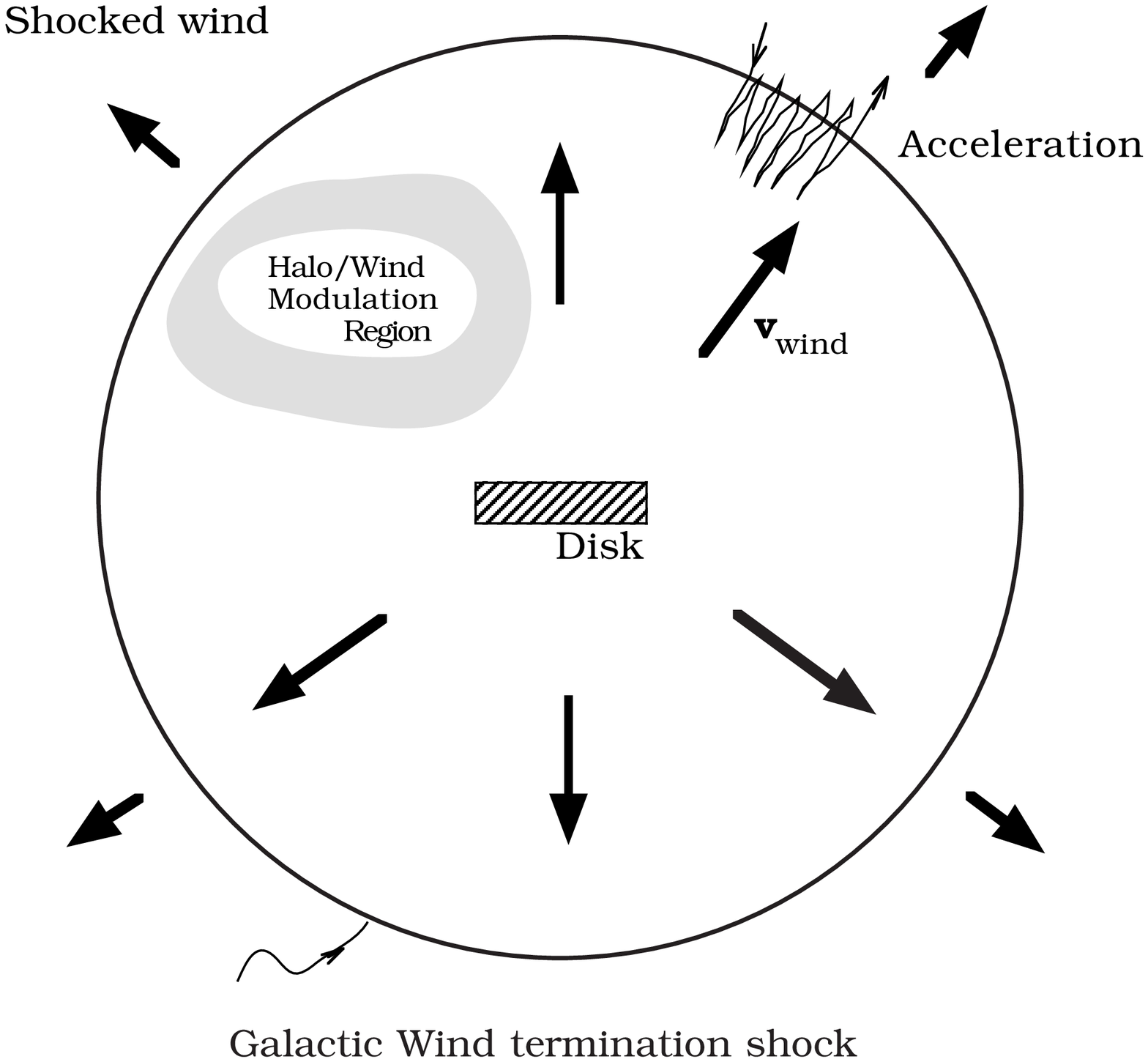}
\includegraphics{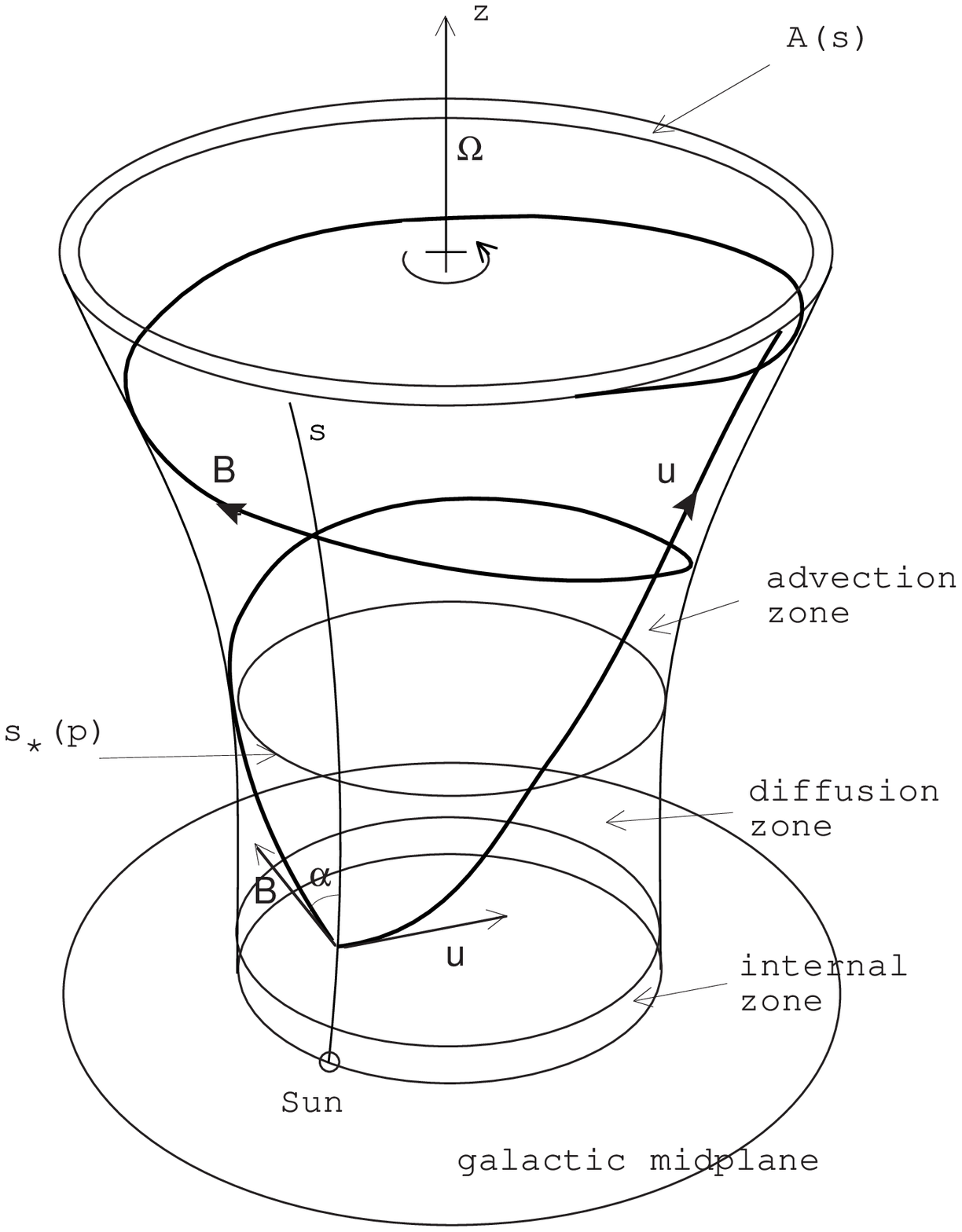}
\caption{ { \bf a}
(left panel). Schematic of particle acceleration at a Galactic Wind termination
shock. In the halo, at distances large compared to the radius of the
galaxy (Disk), the wind velocity becomes radial and constant in magnitude.
It goes through a strong shock transition at a distance where the ram
pressure of the wind has decreased towards the external (IC) pressure.
Particles injected from the heated thermal gas of the shocked wind get
diffusively accelerated at this shock with high efficiency.
{\bf b} (right panel). Schematic of the flow and field configuration of an 
axisymmetric Galactic Wind from a rotating disk galaxy like the Milky Way, 
cf. Ptuskin et al. (1997). 
The meridional flow velocity {\bf u} from a circle in the
disk (like that corresponding to the Sun's distance from the center of
rotation) is roughly parallel to the axis of rotation to flare out at
large distances. The magnetic field {\bf B} becomes slowly azimuthal at
large meridional distances $s$. 
 }
\end{figure}

Fresh particle acceleration will occur with high efficiency at the strong
wind termination shock, at distances $\sim \! 100 \,\rm kpc$. The wind magnetic
field will still intersect the shock at an angle $\sim\! 10^\circ$
(e.g. Zirakashvili et al. 1996), so that the standard process of
diffusive shock acceleration approximately still works there (Fig.~2a).
Assuming an overall acceleration efficiency of 10 to 30 percent, one gets
a total gas internal energy $E_{\rm gas}^{\rm GW} \sim {\rm few}\,
 10^{62}\,$erg and a nonthermal energy 
 $E_{\rm CR}^{\rm GW} \sim 10^{62}$ erg for a system like the
Perseus cluster, ultimately from star formation and subsequent SN
explosions. Since the galaxies are distributed across the
cluster quasi-uniformly, this will also be true for the {\it nonthermal 
particle population and the radiation} they  emit.
The continuing gravitational contraction/accretion of the cluster will
subsequently energize CRs and thermal gas at least
adiabatically, or shock accelerate/heat both components, so that finally
the total energy $E_{CR}$ of energetic particles reaches $10^{63}\, {\rm
erg}\! \sim 0.1\, E_{\rm gas}$ in the cluster; $E_{\rm gas}$ now denotes 
the total
internal energy of the ICM gas (Paper I). The resulting nonthermal energy
density of some tenths of $\rm eV/cm^{3}$ happens to be
roughly equal to that in the interstellar medium of our Galaxy. 

It is instructive to compare the expected nonthermal energy with the
thermal energy content of the cluster galaxies. Assuming the stars
internally to be in virial equilibrium and, for purposes of estimate, all
of them to have a solar mass and radius, then $E_{\rm th}^{\rm star}\! 
\sim (3/10) GM_{\odot}^{2}/R_{\odot} \simeq 10^{48}\,\rm erg$. For a total mass
of about $10^{14} M_{\odot}$ contained in the {\it galaxies} of the Perseus
cluster this gives a total thermal energy in stars $ \sim\! 10^{62}\,\rm erg$,
and thus $E_{\rm CR} > \sum_{\rm gal}\sum_{\rm stars} E_{\rm th}^{\rm star}$. 
This means that
the nonthermal ICM energy is larger than the total thermal energy of all
the stars in all the galaxies contained in the cluster!

It has been argued more recently that, apart from star formation and
overall gravitational contraction, also individual giant radio galaxies
should have injected large and in fact comparable amounts of nonthermal
particles during the life time of a cluster (En{\ss}lin et al. 1997;
Berezinsky et al. 1997). This is no doubt an important additional
possibility. A weakness of this argument consists in the fact that per se
it is predicated on statistical knowledge about the luminosity function
for active galaxies in clusters in general, and not on direct observations
of the individual cluster to which it is applied.

{\it Particle Confinement}

The large-scale magnetic field in the ICM gas may be quite chaotic and not
well connected over distances exceeding typical intergalactic distances
(see section 2). Thus energetic particles may not readily escape from the
cluster due to such topological characteristics. However, already pure
pitch angle diffusion along magnetic field lines with superposed
turbulent fluctuations gives important insights into the confinement
properties of galaxy clusters. Standard quasilinear theory yields a
spatial diffusion coefficient $\kappa_{\parallel}$ along the large scale
field $\bf B$ due to a power spectrum $P(k)$ of magnetic field
fluctuations with wavelength $\lambda=2 \pi /k$ as
$$\kappa_{\parallel}= (1/3) c r_{\rm g}(p) \frac{{\bf B}^2}{\int_k^{\infty}
{\rm d}k' P(k')}, $$
where $p$ denotes particle momentum, $kr_{\rm g}(p)\simeq 1$, and $k$
denotes wavenumber of the field fluctuations.
Let us assume a relative fluctuation field strength of order unity at the
inter-galaxy distance $1/k_{0}$, i.e. a totally turbulent field $P(k_0) \times
k_0 \sim {\bf B}^2$ on this scale, and a power law form of
$P(k)=P(k_0) (k/k_0)^{-n}$. Then 
the diffusion time across the cluster
$T_{\rm esc} \sim R_{\rm cl}^2 / \kappa_{\parallel} > T_{H}$
for $(cp)_{\rm protons} \leq 10^{17}\, {\rm eV \; and} \leq10^{15}\,{\rm eV}$, 
for $n=3/2$ and $n=5/3$, respectively (Paper I).
Also $t_{\rm loss}^{\rm protons}\gg T_{H}$ for nuclear collisions in the 
ICM gas.  
Therefore (except at subrelativistic energies with their prevailing
Coulomb losses), up to these energies CR hadrons accumulate in the cluster
since the galaxy formation epoch, and that is what we called {\it
cosmological} CRs before. The situation is different for relativistic
electrons, which suffer radiative losses:
energetic electrons observed now must either be secondaries or be rather 
recently accelerated.

\section{ Intracluster Magnetic Fields}

The large magnetic field strengths of $B \sim 1 \,\mu \rm G$ in the IC
medium of rich clusters, in particular as observed by Faraday rotation
measurements (e.g. Kronberg 1994), are not easily explained by a
contemporary mechanism because present day turbulent dynamo effects in
such a large-scale system should be extremely slow. Therefore we suggest
here a field configuration that is due to the early formation history of
galaxy clusters and that should be essentially preserved to this day. It
should even be still in a state of development at the present epoch. The
argument derives from the violent early Galactic Winds which accompany the
starbursts responsible for the predominance of the early type galaxies in
rich clusters.

We assume first of all that the protospirals, whose mergers constitute the
building blocks for the E and S0 galaxies, had already generated galactic
magnetic fields of $\mu$G strength. This should indeed be possible within
about $10^8$ yr, i.e. of the order of a rotation period of our Galaxy,
from turbulent dynamo action that invokes boyancy effects from CRs and
magnetic reconnection on spatial scales of $O(100 {\rm pc})$ (Parker,
1992); such a time scale and the resulting field strengths correspond to
generally accepted numbers. In the second stage the ensuing Galactic Winds
extend these fields from the interacting galaxies to almost intergalactic
distances. In the final and by far longest stage, that lasts until now, the
fields are recompressed by the contraction of the cluster to its present
size.

The ICM fields do not reconnect on the intergalactic scale in a Hubble
time. Consequently there is no need for a continuous regeneration of these
fields since their formation. However, this also implies that a
topologically connected overall ICM field will on average not be formed
either, and that the ICM field is chaotic on a scale smaller or equal to
the present intergalactic distance.

In detail we draw on arguments we have in the past used for the field
configuration in a Galactic Wind from our own Galaxy (Zirakashvili et al.  
1996; Ptuskin et al. 1997). They are based on estimates of the relative
amount of field line reconnection vs. the extension of galactic field
lines by a wind to "infinity" (Breitschwerdt et al. 1993). The basic
result was that the rates of reconnection - and thus of the formation of
"Parker bubbles" leaving the galaxy by their boyancy and allowing the
generation of the disk magnetic field - and of extension of this field into
the galactic Halo by the pressure forces of the wind are roughly equal.
Thus both effects occur, and in the cluster galaxies, magnetic energy can
be generated on the large scale of the wind at the expense of the thermal
and nonthermal enthalpies produced in the starburst. The geometry of the
field should roughly correspond to straight field lines out to radial
distances $r$ of the order of the starburst (SB) radius, $r_{\rm gal}^{\rm SB}
\sim 1$ kpc, and spherically diverging field lines beyond that. 
The slow rotation of
the system should then lead to an azimuthal field component $\propto 1/r$
which dominates at large distances over any radial component. However, in
contrast to the familiar situation in the Solar Wind equatorial
plane, the axis of rotation is rather parallel than perpendicular to the
flow at the base of the wind, and thus the dominance of the azimuthal
field component is by no means as drastic as in the case of a stellar wind
(Fig. 3).
	
Assuming the ICM pressure $p_{\rm cl}(z=2)$ at the formation stage of the
early type galaxies to be roughly a factor of $10^{-2}$ smaller than it is
at present (after gravitational compression of the ICM gas), the
termination shock distance $r_{\rm sh}$ is found from 
$\rho(r_{\rm sh})\, u^2 /k_{B} \sim 10^{-2} p_{\rm cl}(z=0)/k_{B} \sim 10^2
\,\rm K/cm^{-3}$. 

For M82 analogs, but with $r_{\rm gal}^{\rm SB} \sim 1$ kpc, we have
$p_{\rm gas}/k_{B} \sim 10^7 \,{\rm K}{\rm cm}^{-3}$. Thus
$r_{\rm sh}/r_{\rm gal}^{\rm SB} \sim p_{\rm gas}/
[10^{-2}p_{\rm cl}(0)] \simeq 300$,  
and therefore
$$
\frac{r_{\rm sh}}{d_{\rm gal}^{\rm field}(0)/(1+z)} \simeq 
\frac{300 \,{\rm kpc}}{(2\,{\rm Mpc}/3)} \simeq 0.5
$$
The Wind Bubble containing the shock-heated wind gas will have a radius
still exceeding $r_{\rm sh}$. Thus, even though the volume of hydrodynamically
unaffected ICM gas may be large enough so that the ICM gas mass exceeds the
mass associated with galaxies by a factor of a few - as observed - there may
be that some Wind Bubbles touch. However, at the scale of $r_{\rm sh}$,
reconnection with a speed between 1 and 10 percent of the Alfv$\acute{e}$n
velocity is too slow to occur over a Hubble time, even in a present-day ICM
magnetic field as high as $10^{-6}$ G. Therefore, on average, the field
structure should remain topologically disconnected until today.

The field strength $B_{\rm cl}(z=2)$ in the Wind Bubbles 
should be of the order of
$$
B_{\rm cl}(z=2) \simeq 4 B_{\rm gal} r_{\rm gal}^{\rm SB}/r_{\rm sh} \sim
10^{-2}B_{\rm gal} \sim 10^{-8} {\rm G}
$$
or somewhat larger, if the field in the bubble increases in the
decelerating postshock flow.
The ongoing cluster contraction/accretion compresses the field to lowest
order isotropically $\propto l^2$, with the scale factor
$$
l \simeq [n_{\rm cl}(0)/n_{\rm bar}(0)]^{1/3}/(1+z),
$$
where $n_{\rm cl}(0)\sim (10^{-3}\,{\rm to}\,
10^{-4})\,{\rm cm}^{-3}$ and $z=2$.
Choosing for the present mean baryon number density the value
$n_{\rm bar} \sim 3 \times 10^{-7}\,{\rm cm}^{-3}$, we obtain $l \sim 
(2.3\,\rm to\, 5)$. 

Finally then, we obtain for the present-day ICM field:
$B_{\rm cl}(z=0)/B_{\rm cl}(z=2) \sim (5 \,\rm to \,25)$. 
Therefore the present-day ICM magnetic field should have a mean strength of
the order of $10^{-7}$G, from "primordial" seed fields, randomly directed
on an intergalactic scale. Although smaller by about one order of
magnitude than estimated from Farady rotation measurements, such fields
need not necessarily be unrealistic, considering that observations might
emphasize regions of high magnetic fields. In addition the increase of the
field in the Galactic Wind bubbles beyond their postshock value might be
more than a factor of unity as assumed above. Thus we cannot exclude
$\mu$G fields although they certainly are at the upper limit our estimate
permits. The interpretation of recent UV data also points to 
small field values (see section 3).

In conclusion, there is hardly any need for a contemporary "turbulent IC
dynamo". Nevertheless, the estimated CR enthalpy of $\sim 0.5 \,{\rm
erg/cm^3}$ is essentially a free energy reservoir for a future 
increase in $B$ towards a mean strength of a few $\mu$G.
In the next section we consider the nonthermal radiation from clusters.
In particular we shall discuss the implications of the broad-band 
observations for the nonthermal energy content and ICM magnetic fields.

\section{Nonthermal emission}

From the Coma cluster radio fluxes are measured from 10.3\,MHz to 2.7\,GHz
(Bridle and Purton, 1968; Henning 1989; Kim et al. 1990; Giovannini et al.
1993), and at 4.85\,GHz an upper flux limit was reported by Kim et al.
(1990). The energy fluxes $J_{\nu} \propto \nu^{-\alpha_{\rm r}}$ between
30.9 MHz and 1.4 GHz are well fitted with a power-law index $\alpha_{\rm
r} =1.16$ (Bowyer \& Bergh\"ofer 1998).  The data at 2.7 and 4.85 GHz fall
below this extrapolation, which may indicate a steepening in the electron
spectrum (Schlickeiser et al. 1987) for Lorentz-factors $\gamma\geq
5\times 10^4$, but can be explained also as an instrumental effect (Deiss
1997). Below we use an index $\alpha_{\rm r} =1.16$.

In the extreme ultraviolet (EUV) region, diffuse radiation between 65 and
245 eV is observed which appears in excess of the thermal fluxes of the
X-ray emitting gas with $T\sim 2\times 10^6\,\rm K$, and 
this radiation was initially interpreted in terms of thermal emission
from a gas of lower temperature, $T\sim 8\times 10^5\,\rm K$ (Lieu et al.
1996). Subsequently, the excess EUV emission, observed in a number of
clusters (e.g. Mittaz et al. 1998), was suggested to represent inverse
Compton (IC) radiation (Hwang 1997; Sarazin and Lieu 1998) of low energy
electrons, $\gamma\sim 300$, on the 2.7 K microwave background radiation.
Also En{\ss}lin and Biermann (1998) used this possibility to estimate lower
limits to the magnetic field strength in the Coma cluster (see also Lieu
et al. 1998).

Bowyer and Bergh\"ofer (1998) have shown that the size of diffuse emission
of the Coma cluster is significantly larger in radio than in EUV
light. In addition, the spectral index of the energy flux $J(E)$ of the
EUV emission is $\alpha_{\rm uv}\approx 0.75$ which is significantly
different from that in the radio domain. Based on these arguments, Bowyer
and Bergh\"ofer (1998) suggested that two different populations of
relativistic electrons should be responsible for the radio synchrotron and
the EUV IC fluxes. We shall show below that the differences in these
spectral indices are quite naturally explained in terms of a single
population of electrons producing both the EUV excess IC and the radio
synchrotron emission which avoids the need for a 2-component model.

The characteristic synchrotron frequency produced by electrons with
Lorentz factor $\gamma$ in the magnetic field $B$ is $\nu \simeq (B/1\,\mu
{\rm G}) \gamma^2\,\rm Hz$ (e.g. Ginzburg 1979).  Thus in a magnetic field
of $B\sim 10^{-7}\,\rm G$ the production of synchrotron radiation with
$\alpha_{\rm r} =1.16$ in the region from 30 MHz to 1.4 GHz requires the
energy distribution of the electrons $N(\gamma)\propto
\gamma^{-\alpha_{\rm e}}$ to extend with an index $\alpha_{\rm
e}=1+2\alpha_{\rm r}\approx 3.3$ from $\gamma_1 \simeq 1.7\times 10^4$ to
$\gamma_2 \simeq 1.4\times 10^5$. These Lorentz factors are smaller by a
factor of 3 if $B\sim 10^{-6}\,\rm G$.

For the IC process the mean energy of photons is $\epsilon \simeq (4/3)
\epsilon_0 \,\gamma^2$, where $\epsilon_0 $ is the energy of target 
photons. For the 2.7\,K background radiation $\epsilon_0 = 6.5\times
10^{-4}\,\rm eV$. For the production of Coma's EUV spectrum from 65 to 250
eV one therefore needs a spectral index of electrons $\alpha_{\rm e}\simeq
2.5$ only in a rather narrow energy region around $\gamma_{\rm ic} \simeq
300$. This energy is much smaller than the energy $\gamma_1$ necessary for
the radio emitting electrons. Thus a scenario in which radiative losses
induce a break in $\alpha_{\rm e}$ for a single population of electrons
somewhere in between $\gamma_{\rm ic}$ and $\gamma_1$ can readily account
for the observed indices both in the radio and the EUV. Indeed, assuming
that the spectral index of the source function of electrons in the ICM is
$\alpha_{\rm inj}=2.3$, the IC emission of those electrons at low energies
will give $\alpha_{\rm uv}=0.65$, which is rather close to the value 0.75,
in a rather narrow EUV band. After the break,
the electron distribution steepens to $\alpha_{\rm e}=1+\alpha_{\rm
inj}=3.3$, required for the radio emission of Coma.

\setcounter{figure}{2}
\begin{figure}
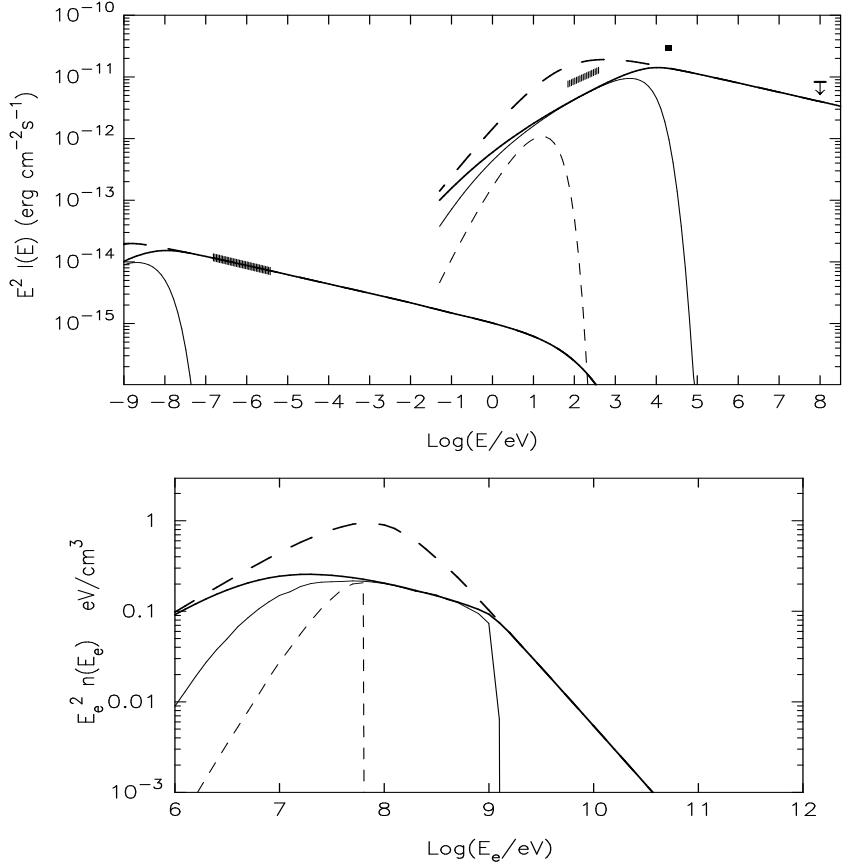

\vspace{11.5cm}
\includegraphics{Fig4.ps}
\includegraphics{Fig5.ps}
\caption{Synchrotron and IC fluxes (top panel)
and the energy distribution of electrons (bottom panel)
 calculated assuming continuous (heavy lines) and impulsive (thin lines)
injection of electrons with $\alpha_{\rm inj} =2.3$ and
$\gamma_{\rm c}=2\times 10^8$, for the
injection times  $t=10^9\,\rm yr$ (solid) and
$t=10^{10}\,\rm yr$ (dashed). The ICM gas density 
$n=10^{-3} \,\rm cm^{-3}$. For the assumed  
 $B=0.1\,\rm \mu G$ the electrons are stationarily 
injected with 
$L_{\rm inj}=2.6 \times 10^{45} \,\rm erg/s$; the total
energy input during 1\,Gyr is $8.2\times 10^{62}\,\rm erg$.
For the impulsive injection, the same energy inputs as in the corresponding
 cases of continuous injection are assumed.  
Nonthermal fluxes from Coma in the radio, EUV, X-ray, and the EGRET 
upper flux limit (Sreekumar et al. 1996) are also shown. 
 }
\end{figure}

Such a break in the energy spectrum is necessarily produced by radiative
losses of electrons on time scales $\geq 10^9\,\rm yr$. Indeed, the
characteristic energy loss time in the 2.7\,K background radiation field
as well as due to synchrotron emission can be written as 
\begin{equation}
t_{\rm rad}=2.4\times 10^{12} [C(\gamma)+0.1 B_{\mu \rm G}^{2}]^{-1} 
\,\gamma^{-1} \;  \rm yr \; , 
\end{equation} 
where $B_{\mu \rm G}=B/1\,\mu \rm G$.
The coefficient $C(\gamma)$ takes into account corrections for the
Klein-Nishina effect in the IC energy losses of ultrarelativistic
electrons, which is important for $B\leq 3\,\rm \mu G$.
$C(\gamma)=1$
for $\gamma\ll 10^8$ (IC losses in the Thompson limit), but 
$C(10^8)\simeq 1/2$, $C(2\times 10^8)\simeq 1/3$, and 
$C(5\times 10^8)\approx 1/6$.

For a magnetic field $\leq 3 \,\mu \rm G$ and timescales $\simeq
(1-3)\times 10^{9} \,\rm yr$, equation (1) predicts the radiative cutoff
energy $\gamma_{\rm br} \sim 10^3 $. It is important to note that
relativistic electrons need to be produced (accelerated) in the ICM {\it
continuously} during all these last years, because in the case of an
`impulsive' injection of the electrons the radiative losses would remove
all particles with energies above $\gamma_{\rm br}$, and then the radio
spectra cannot be explained.

This is seen in Fig.\,3 which shows the results of calculations that
assume both continuous and impulsive injection of electrons with a
spectrum
\begin{equation}
Q(\gamma)\propto \gamma^{-\alpha_{\rm inj}} \exp (-\gamma/\gamma_{\rm c}),
\end{equation}
where $\gamma_{\rm c}$ defines the assumed characteristic maximum energies
of accelerated particles. The fluxes for continuous injection are
normalized to the radio flux $2\,\rm Jy$ observed at $400\,\rm MHz$. The
IC radiation for continuous injection during the last
$t=10^{9}\,\rm yrs$ (heavy solid line)  has the spectral shape of the EUV
flux, and for a slightly smaller magnetic field (cf. Fig.4), or an
injection time larger by a factor of 2, it can also explain the absolute
EUV flux. However in the case of relativistic electron production on much
larger time scales, $\sim 10^{10}\,\rm yr$, the explanation of the shape
of the EUV radiation becomes problematic even for continuos injection.
Moreover, the agreement with observed spectra becomes worse than shown in
Fig.3 if we take into account that for cosmological timescales the energy
density of the microwave background increases with redshift as $w_{\rm
mbr}\propto T^{4}(z)\propto (1+z)^4$. Thus, the electrons
responsible for the excess EUV radiation, if indeed it has an IC origin,
must have been produced continuously during the recent (1-3)\,Gyrs.  This
would be the characteristic age of the accretion shocks in the cluster which
seem to be the most probable accelerators for the radio emitting
electrons.

An IC origin of the excess EUV flux imposes a strong constraint on the
magnetic field in Coma. The variation of the IC flux for different
magnetic fields are shown in Fig.4 which demonstrates that the lowest
consistent magnetic field strength in Coma is equal to $B=7.5\times
10^{-8}\,\rm G$. This field could explain also the excess flux observed by
{\it Beppo-SAX} in hard X-rays beyond 25 keV (Fusco-Femiano et al. 1998), that
might be due to IC radiation of GeV electrons. The dot-dashed curve in
Fig. 4 falls below both the X-ray and the EUV data. However, the assumption
of
injection for about 5 \,Gyr shifts the position of the radiative cutoff
energy by factor of 5 and increases by the same factor the number of
low energy electrons. The number of electrons with
$\gamma\sim 300$ is also increased for a steeper 
injection spectrum, $\alpha_{\rm inj} = 2.5$.  Both these options could
marginally explain the radio synchrotron and EUV IC fluxes.
Thus an IC origin of the excess EUV radiation restricts the cluster
magnetic
field to a narrow range around $0.1\,\rm \mu G$.

\begin{figure} \vspace{5.8cm} \includegraphics{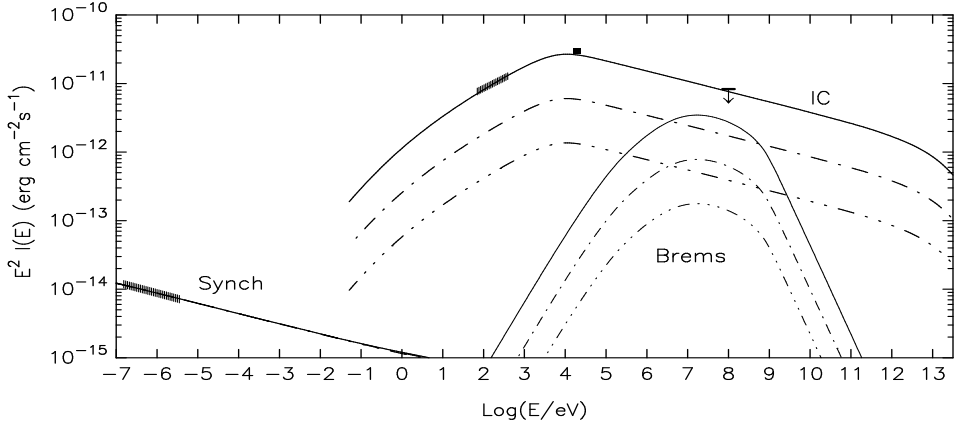}
\caption{Nonthermal radiation fluxes calculated for 
continuous injection of electrons during $t= 10^9\,\rm yr$, for different
ICM magnetic fields: $B= 7.5\times 10^{-8}\,\rm G$ (solid lines),
$1.5\times 10^{-7}\,\rm G$ (dot-dashed lines), and $3 \times 10^{-7}\,\rm G$ 
(3-dot--dashed lines)}
\end{figure}

For magnetic fields $B\simeq 10^{-7}\,\rm G$ the observed radio fluxes
require injection of accelerated particles with a luminosity $L_{\rm
inj}\simeq 3\times 10^{45}\,\rm erg/s$. In a thermal gas with density
$n\sim 10^{-3}\,\rm cm^{-3}$, a {\it secondary} origin of the
relativistic electrons, i.e. their production in $pp$ interactions, would
require an enormously large total energy in relativistic protons. Indeed,
the luminosity in the $\pi^{\pm}-\mu^{\pm}$-decay electrons can be
estimated as
\begin{equation} L_{\pm}\simeq 7.7\times 10^{40}\,(n_{\rm
p}/10^{-3}\,{\rm cm^{-3}})^{-3} (E_{\rm CR}/10^{60}\,\rm erg)\; erg/s.
\end{equation}

The fluxes of IC gamma-rays to be expected at TeV energies are shown in
Fig.\,5. for the case of $B=10^{-7}\,\rm G$ and acceleration of
electrons up to an exponential cutoff energy $E_{\rm
c}=100\,\rm TeV$. For magnetic fields
$\sim 0.1\,\rm \mu G$ the diffusive shock acceleration mechanism in the
Bohm limit still allows such high values for accelerated electrons. In
this limit the acceleration time, defined by the rate
${\rm d}\gamma/{\rm d}t \simeq \gamma u^2/D(\gamma)$ where $u$ is the
shock speed and $D(\gamma)$ is the diffusion coefficient, is estimated as
\begin{equation}
t_{\rm acc}\simeq 5.4\times 10^{-5} \,u_{3}^{-2} B_{\rm
\mu G}^{-1} \,\gamma \;\rm yr\;,
\end{equation}
where $u_3 = u/10^3\,\rm km/s$. Equating $t_{\rm acc}$ with $t_{\rm rad}$
shows that values $\gamma\geq 10^8$ are possible.

\begin{figure}
\vspace{4.5cm}
\includegraphics{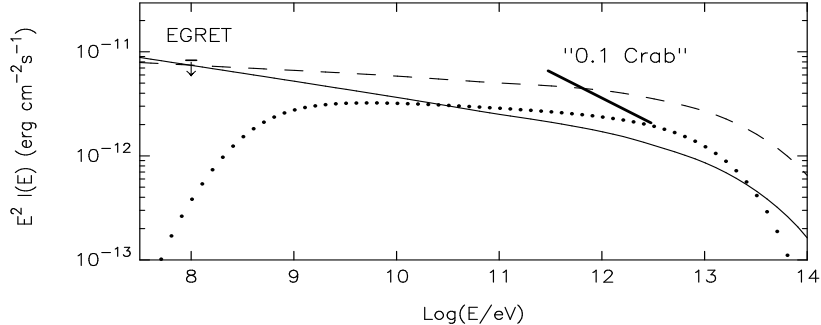}
\caption{IC $\gamma$-ray fluxes expected from the Coma 
for $B =7.5\times 10^{-8}\,\rm G$ and $\alpha_{\rm inj} =2.3$ (solid lines), 
and $B =9\times 10^{-8}\,\rm G$ and $\alpha_{\rm inj} =2.1$ (dashed lines). 
The dots show the $\gamma$-ray fluxes 
produced in $pp$ interactions of CRs with $E_{\rm CR}=
3\times 10^{62}\,\rm erg$ in ICM with $n=10^{-3}\,\rm cm^{-3}$.  
}
\end{figure}

The solid line in Fig.\,5 corresponds to the solid curve in Fig.\,3.  The TeV
fluxes of IC $\gamma$-rays in that case correspond to several per cent of
the TeV emission from the Crab Nebula. In the case of single power law
injection with $\alpha_{\rm inj} =2.3$ a further increase of the IC fluxes
at TeV energies are impossible because of the EGRET upper limit above
 100 MeV (Sreekumar et al 1996). Note, however, that for a more realistic
 modelling of Coma as a spatially nonuniform source, the power law index
 $\alpha_{\rm inj}$ could be somewhat smaller than 2.3. We do not consider
 here this possibility, which should address also the question of different
 angular sizes of Coma in EUV and radio, as well as the flattening of the
 radio spectra observed towards the core of the cluster (Giovannini et al.
 1993; Deiss et al. 1997). In Fig.\,5 the dashed line shows the maximum of
 the IC radiation fluxes to be expected in the case of $\alpha_{\rm inj}
 =2.1$, which is then at the 0.1 Crab level. The dots correspond to the
 $\pi^0$-decay fluxes produced by hadronic CRs with $\alpha_{\rm cr}=2.1$,
 assuming $E_{\rm CR}=3\times 10^{62}\,\rm erg$. The size of the radio
 emission produced by high-energy electrons in Coma is about half a degree.
 Therefore we can expect in that case a similarly large size for the TeV
 emission. Detection of such extended and weak ($\leq\,$0.1 Crab) TeV
 fluxes by current instruments is problematic, but they are quite
 accessible for the future HESS and VERITAS arrays.

These results for the IC fluxes relate to the case of ICM magnetic fields
$B\simeq 10^{-7}\,\rm G$. However, the estimates for the magnetic fields
in Coma range from 0.1 to several $\mu \rm G$. In particular, Kim et al.
(1990) and Feretti et al. (1995) deduced magnetic fields $B\simeq 1.7\,\rm
\mu G$ and $B\simeq 6.0\,\rm \mu G$, respectively, from Faraday rotation
measurements on background radio sources. If this is so, then an IC origin
of the excess EUV fluxes is absolutely excluded. The only other
possibility left for an explanation of this radiation in terms of
nonthermal radiation is synchrotron production by very high energy
electrons. The interpretation of the steep radio fluxes then requires a
second component of nonthermal electrons.

For a production of synchrotron photons with energies 200\,eV ($\nu \sim
5\times 10^{16}\,\rm Hz$) in magnetic fields $B\sim 1\,\rm \mu G$ one
needs electrons with $\gamma \geq 2\times 10^8$. From Eq.(1), the
radiative loss times of these electrons is very short, $t_{\rm rad}\leq
10^4 \,\rm yr$. Even assuming a rectilinear propagation of such electrons,
the maximum possible distance from the acceleration sites would be less
than 3\,kpc. Thus, the acceleration sites of the electrons should be
rather smoothly distributed in the ICM in order to result in a smooth
distribution of the synchrotron EUV radiation.  This seems quite
reasonable for accretion shocks, but not for isolated point sources
like active galaxies. Another requirement is that the spectrum of accelerated
particles should be very hard, with $\alpha_{\rm inj}\simeq 1.5$. Then the
spectrum of electrons steepens to $\alpha_{\rm e} \simeq 2.5$ which is
required for the production of synchrotron radiation with $\alpha_{\rm uv}
\simeq 0.75$.

In Fig.\,6 we show the spectra of the synchrotron and IC radiation generated
in a magnetic field $B=2\,\rm \mu G$ by two different populations of
relativistic electrons. The dashed curves are produced, as in previous
figures, by electrons continuously injected into the ICM during recent
epochs up to $t\geq 10^9\,\rm yr$ with $\alpha_{\rm inj}=2.3$ and
luminosity $L_{\rm inj}=3.5\times 10^{42}\,\rm erg/s$.  For this
luminosity, these electrons can be well explained as {\it secondaries}
produced in the interactions of hadronic CRs, of total energy $E_{\rm
CR}=4.5\times 10^{61}\,\rm erg/s$, with an ICM gas of density $n_{\rm
p}=10^{-3}\,\rm cm^{-3}$. They can generate the observed radio emission.
On the other hand, the EUV fluxes in Figs. 3 and 4 are produced by the
second component with $\gamma_{\rm c}=6\times 10^8$, and a hard injection
spectrum with $\alpha_{\rm inj}=1$ (solid and dot-dashed curves) and
$\alpha_{\rm inj}=1.5$ (three-dot--dashed). The curves show that
continuous injection of this second component over times exceeding
$3\times 10^7\,\rm yr$ from now, without reacceleration, would lead to an
accumulation of electrons at energies smaller than $\gamma\sim10^5$. This
would be quite sufficient to produce a radio synchrotron flux above the 
observed
level. Moreover, in the case of $\alpha_{\rm inj}=1.5$ one has to assume a
cutoff in the injection spectrum below $\gamma_{\rm low}\simeq
10^5$. Otherwise the number of low energy electrons in the injection
spectrum would be unacceptably high.

\begin{figure}
\vspace{6.5cm}
\includegraphics{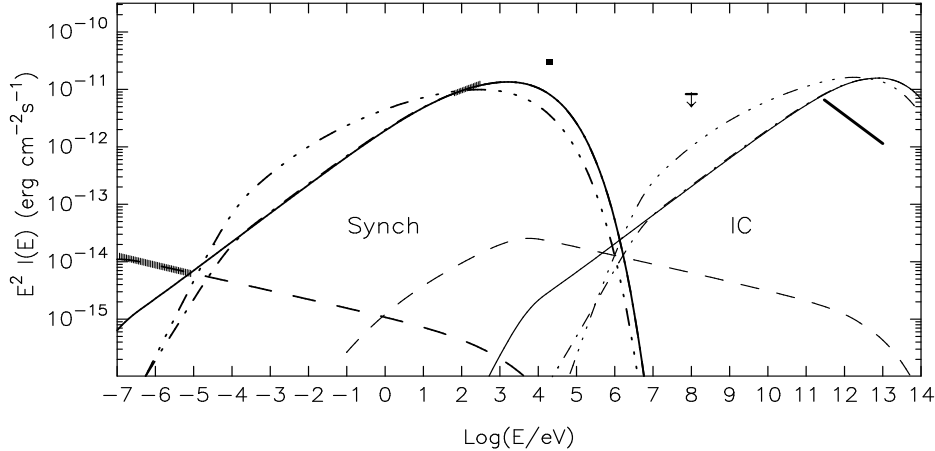}
\caption{Nonthermal fluxes expected in the case of $B =2\,\mu
\rm G$ in a two-component model for relativistic electrons in Coma. For
the first (radio) component (dashed lines), continuous injection during
$t=10^9\,\rm yr$ is assumed, with a spectrum of the form of Eq.(2) with
$\alpha_{\rm inj} =2.3$. For the second (EUV) component the parameters of
the continuous injection, without reacceleration of electrons, read:
$t=10^{9}\,\rm yr$ and $\alpha_{\rm inj}=1$ (solid lines), $t=3\times
10^{7}\,\rm yr$ and $\alpha_{\rm inj}=1$ (dot-dashed), $t=3\times
10^{7}\,\rm yr$ and $\alpha_{\rm inj}=1.5$ (three-dot-dashed). For the
case of $\alpha_{\rm inj}=1.5$ a rapid transition to a $\propto \gamma^2$
behavior in the injection spectrum in the region $\gamma\leq 10^5$ is
assumed. The exponential cutoff energy is $\gamma_{\rm c}=6\times 10^8$,
and the gas
density is $n=10^{-3}\,\rm cm^{-3}$. Injection rates for the first
and second components are $L_{\rm inj}= 3.5\times 10^{42}\,\rm erg/s$ and
$L_{\rm inj}\approx 5.5\times 10^{44}\,\rm erg/s$, respectively. The total
number of electrons in the second component is $N_{\rm e}\simeq 10^{58}$. 
The heavy bar corresponds to the "0.1 Crab" flux level.}
\end{figure}

Thus, the scenario with high magnetic fields and nonthermal origin of the
excess EUV radiation requires that the second electron component 
enters a reacceleration cycle at least once in
 $3\times 10^7$ year. In addition, this component should not
contain electrons with low energies. Therefore it cannot be produced by
reacceleration of secondary electrons. 

In a speculative vain this
component could be due to run-away pulsars born with kick velocities
$\sim 1000\,\rm km/s$.
Another possibility for the pulsars to appear in the ICM
would be that they could be dragged there in the process of
galaxy-galaxy collisions. Pulsars produce, at the pulsar wind termination
shocks, an electron distribution with a strong deficit of low-energy
particles (e.g. Arons 1996). These electrons could be kept at those high
energies by entering into frequent reacceleration processes on the
accretion shocks in the ICM, and could be thus  responsible for the
second component.

Reacceleration would not affect the radio electrons if it happened in
the central region (see Bowyer and Bergh\"ofer 1998). The signature of the
second component would be a rising spectrum of IC $\gamma$-rays. Fields
below  $1\,\rm \mu G$ would even imply uncomfortably large TeV fluxes from
the center of the Coma cluster. Thus the alternative between low and high
magnetic fields in the Coma cluster would be given by the shape of the TeV
$\gamma$-ray spectrum.
 \vspace{2mm}
 
\noindent                                         
{\bf Acknowledgements}~~The authors thank F.\,A.\,Aharonian and
E.\,N.\,Parker for illuminating discussions. 
They also thank H.\,B\"ohringer for the permission to use Fig.\,1 here.


\end{document}